\documentclass[12pt]{article}
\usepackage{graphicx}
\newcommand{\beq}{\begin{equation}}
\newcommand{\eeq}{\end{equation}}
\newcommand{\bea}{\begin{eqnarray}}
\newcommand{\eea}{\end{eqnarray}}

\def\fun#1#2{\lower3.6pt\vbox{\baselineskip0pt\lineskip.9pt
  \ialign{$\mathsurround=0pt#1\hfil##\hfil$\crcr#2\crcr\sim\crcr}}}

\begin{document}
\begin{titlepage}
\begin{flushleft}
       \hfill                      {\tt hep-th/0206128}\\
       \hfill                       FIT HE - 02-02 \\
\end{flushleft}
\vspace*{3mm}
\begin{center}
{\bf\LARGE Origin of Small Cosmological Constant \\ }
{\bf\LARGE in Brane-World\\ }
\vspace*{5mm}

\bigskip

{\large Kazuo Ghoroku\footnote{\tt gouroku@dontaku.fit.ac.jp}\\ }
\vspace*{2mm}
{
\large 
${}^2$Fukuoka Institute of Technology, Wajiro, Higashi-ku}\\
{
\large 
Fukuoka 811-0295, Japan\\}
\vspace*{5mm}

{\large Masanobu Yahiro \footnote{\tt yahiro@sci.u-ryukyu.ac.jp} \\}
\vspace{2mm}
{
\large 
${}^4$Department of Physics and Earth Sciences, University of the Ryukyus,
Nishihara-chou, Okinawa 903-0213, Japan \\}

\vspace*{10mm}

\end{center}

\begin{abstract}
We address the relation among the parameters
of accelerating brane-universe embedded
in five dimensional bulk space. It is pointed out that the tiny cosmological
constant of our world can be obtained as quantum corrections
around a given brane-solution in the bulk theory or
in the field theory on the boundary from the holographic viewpoint. 
Some implications to the cosmology and constarints on the 
parameters are also given.

\end{abstract}
\end{titlepage}

\section{Introduction}
It is quite expectable that our four dimensional world can be 
regarded as a brane like the one proposed by 
Randall and Sundrum (RS brane) in ~\cite{RS1,RS2}.
It might be formed in
the process of compactification
from the ten-dimensional superstring theory. The D-brane
approach is getting very useful for the study of such theory.
In particular, there is some interest
 in the geometry obtained from the D3-brane of type IIB theory.
Near the horizon of the stacked D3-branes, 
the configuration  $AdS_5\times S^5$ is realized and 
the string theory on this background  describes
the four-dimensional SUSY Yang-Mills theory which lives on the
boundary of $AdS_5$ \cite{M1,GKP1,W1,Poly1}. This holographic correspondence
of the five dimensional gravity and the field theory on the boundary
has attracted many interest.

It would be very suggestive that
a thin three-brane like RS brane could be
embedded in $AdS_5$ as our world.
This idea gives an alternative to the standard Kaluza-Klein (KK)
compactification via the localization of the zero mode of the graviton 
\cite{RS2}.
Brane approach opened also a new way to the construction of
the hierarchy between four-dimensional Planck mass and
the electro-weak scale, and also for realization of the small
observable cosmological constant
 with lesser fine-tuning \cite{ArHa,KaSc}.

The localization of the graviton on such a brane 
with a cosmological constant has also been confirmed 
when it is embedded in  $AdS_5$ \cite{BDEL,bre} or in $dS_5$ \cite{bre}. 
The theory of the gravity under consideration is  five-dimensional. However, 
it could be considered as a part of 10 dimensional theory with compact $S^5$.
This picture has been supported in the case of flat brane
through the KK mode contribution to the
Newton's law on the brane via AdS/CFT correspondence \cite{DL}.
 
Being stimulated by the recent cosmological observation,
many approaches to the cosmology with a small cosmological constant
have been given from the viewpoint 
of brane universe (for example \cite{Bin,BDEL}).
Up to now many solutions for such brane-world have been given, but
the cosmological constant of our world is given by hand.
However its value should be
determined by some dynamical reason in the bulk theory.
It is a challenging problem to resolve this point.

Here we propose a clue to the resolution of this issue by 
pointing out that the 4d cosmological
constant should be determined by the quantum corrections in the bulk
theory. Those corrections can be regarded as the breaking of
the conformal symmetry of the bulk theory for such a brane-solution with
a tiny cosmological constant. 

In Section 2, we give a set-up of accelerating brane solutions 
obtained previously on the basis of 
a simple ansatz imposed on the bulk metric. 
In section 3, the 4d cosmological constant is estimated by considering the
effective 4d action, which is obtained from the bulk theory. And the
necessity of quantum corrections is discussed.
In section 4, phenomenological constraints on the parameter are given, and
the models of the brane-world are restricted.
In the final section, the summary and speculations are given.

\section{Brane solutions for accelerating universe}

We start from a simple 
five-dimensional gravitational action. It is given in the
Einstein frame as\footnote{
Here we take the following definition, $R_{\nu\lambda\sigma}^{\mu}
=\partial_{\lambda}\Gamma_{\nu\sigma}^{\mu}-\cdots$, 
$R_{\nu\sigma}=R_{\nu\mu\sigma}^{\mu}$ and $\eta_{AB}=$diag$(-1,1,1,1,1)$. 
Five dimensional suffices are denoted by capital Latin and the four
dimensional
one by the Greek ones.
}
\beq
    S_5 = {1\over 2\kappa^2}\Bigg\{
      \int d^5X\sqrt{-G} (R -  2\Lambda + \cdots)
          +2\int d^4x\sqrt{-g}K\Bigg\}, \label{action}
\eeq
where the dots denote the matter part,
$K$  being the extrinsic curvature on the boundary. This term is 
necessary to construct the effective 4-dimensional brane action as
shown in the next section, but
it plays no role in solving 5d Einstein equation here.
The fields represented by the dots
are not needed to construct the background. 
The other ingredient is the brane action,
\beq
    S_{\rm b} = -{\tau}\int d^4x\sqrt{-g}, \label{baction}
\eeq
which is added to $S_5$. And the Einstein equation is written as
\beq
 R_{MN}-{1\over 2}g_{MN}R=\kappa^2T_{MN} \label{equation}
\eeq
where $\kappa^2T_{MN}=-(\Lambda+{1\over b}\delta(y)\kappa^2\tau
\delta_{\mu}^M\delta_{\nu}^N) g_{MN}$ and $b=\sqrt{-g}/\sqrt{-G}$.

\vspace{.3cm}
We solve the equation (\ref{equation}) in the following
Friedmann-Robertson-Walker type (FRW) metric,
\beq
 ds^2= A^2(y)\left\{-dt^2+a^2(t)\gamma_{ij}(x^i)dx^{i}dx^{j}\right\}
           +dy^2  \, \label{metrica},
\eeq
where the coordinates parallel to the brane are denoted by $x^{\mu}=(t,x^i)$,
 $y$ being the coordinate transverse to the brane. The position of the brane
is taken at $y=0$. In this case, the three-dimensional metric $\gamma_{ij}$
is described in Cartesian coordinates as,
$
  \gamma_{ij}=(1+k\delta_{mn}x^mx^n/4)^{-2}\delta_{ij}  
$, 
where the parameter values $k=0, 1, -1$ correspond to a
flat, closed, or open universe respectively.
\vspace{.5cm}


For the metric (\ref{metrica}), we obtain from Eq.(\ref{equation}) the following reduced equations \cite{bre}:
\beq
  ({\dot{a_0}\over a_0})^2=\lambda-{k\over a_0^2}, 
   \qquad A'^2+{\Lambda\over 6}A^2=\lambda ,
            \label{Einstein2}
\eeq
where we set $A(0)=1$, but this normalization 
does not affect the generality of our analysis.
Then $\lambda$ is given by
\beq
    \lambda = \kappa^4\tau^2/36+\Lambda/6 . \label{4cos1}
\eeq 
The first equation of (\ref{Einstein2}) is solved for each $k$, but they 
are abbreviated 
since we do not use them here. The solutions for $\lambda>0$ 
of the second equation are obtained under the following boundary condition, 
\beq
  {A'(0+)-A'(0-)}=-{\kappa^2\tau\over 3}. \label{bound2}
\eeq
In the following, we give the solutions for 
$\lambda>0$ since they are used hereafter. Such solutions are obtained
for both cases of $\Lambda>0$ and $\Lambda<0$.

For $\Lambda <0$, $A(y)$ is solved as
\beq
  A(y)= {\sqrt{\lambda}\over\mu} \sinh[\mu(y_H-|y|)] \, \label{metrica4}
\eeq
where 
  $\mu=\sqrt{-\Lambda/6}$, 
$\sinh(\mu y_H)=\mu/\sqrt{\lambda}$ and
$y_H$ represents the position of the horizon in $AdS_5$.
This solution represents a brane at $y=0$. The configuration is taken
to be $Z_2$ symmetric with respect to the reflection, $y\to -y$. 

\vspace{.5cm}

When  $\Lambda$ is positive, the solution for  
 $a_0(t)$ is the same as above, but  
 $A(y)$ is different.
One has
\beq
 A(y)={\sqrt{\lambda}\over \mu_d}\sin[\mu_d(y_H-|y|)].
          \label{desit}
\eeq 
Here 
  $\mu_d=\sqrt{\Lambda/6}$, 
  $\sin(\mu_d y_H)=\mu_d/\sqrt{\lambda}$
and $y_H$ represents the position of the horizon in the bulk $dS_5$, where
there is no spatial boundary as in $AdS_5$.
This configuration represents a brane with $dS_4$ embedded in the
bulk $dS_5$ at $y=0$. The $Z_2$ symmetry is also imposed. 

\vspace{.5cm}

\section{4d Cosmological Constant}

The solutions given above are obtained by solving  5-dimensional
Einstein equation, and the parameters of the theory have no restriction.
In other words, 
a favorite solution is realized by imposing an appropriate
relation on the parameters by hand.
For example, we can set the relation, 
$\kappa^2\tau/6=\sqrt{-\Lambda/6}$, to obtain the original Randall-Sundrum
brane solution, and this relation is known as the fine tuning. Up to now,
no one has given any satisfactory explanation to this relation in terms of some
dynamical reason or symmetries.

On the other hand, it is well known that
the solutions are severely restricted when the
theory has some symmetry and this symmetry should be preserved for the
solutions obtained. In this sense, 
the above relation for the RS brane solution might be considered as
a result of such a symmetry. In fact, we can see 
below that this relation can be 
regarded as a reflection of conformal symmetry
in the bulk theory or in the field theory on the boundary
in the sense of AdS/CFT correspondence. 
Then, we could get a solution with a finite $\lambda$
when this symmetry is broken.

The understanding is confirmed as follows. 
The four dimensional part of the bulk five dimensional equations 
(\ref{Einstein2})
should also be obtained in the same form
from the effective 4d brane action ($S^{\rm eff}_b$). 
And this action $S^{\rm eff}_b$ can be obtained from
the bulk theory via path integral. Although 
the five dimensional theory is our starting point, it would be natural to 
consider this is also derived from ten dimensional superstring theory.
However, we start from the five dimensional theory for simplicity.
Then we can write $S^{\rm eff}_b$ as,
\beq
 S^{\rm eff}_b={1\over 2}S_b+ \ln Z_5(g)
\eeq
\beq
  Z_5(g)=\int_{G|_{y=0}=g} DG D\psi e^{iS_5},   \label{effective}
\eeq
where $S_b$ and $S_5$ are given in the previous section. 
The other fields contained in $S_5$ are denoted by $\psi$.
We notice here (i) the path integration in (\ref{effective}) is performed
for the field defined in the region $y\geq 0$, half of the whole region.
This is justified by the $Z_2$ symmetry of the solution. (ii) The bulk
fields take their boundary values on the brane at $y=0$. (iii) In the
sense of (i), a half of the brane action is considered.

\vspace{.3cm}
In the case of $\Lambda<0$, the bulk space has a boundary where
a dual field theory can be considered in the context of AdS/CFT 
holographic correspondence. 
But this holography would not be appropriate here
since the $AdS_5$
is deformed by a small cosmological constant $\lambda$ and
there is no evidence of the conformal invariance in this case.
So, we should consider some quantum field theory (QFT), in which the conformal
invariance would be slightly broken, instead of
a CFT on the boundary. Further we must introduce counter terms 
\cite{HeSk,BaKr}, 
$S_{\rm CT}$, which are needed to regularize the action
for a class of classical solutions like the given here in the bulk.
Then it is possible to write as \cite{Gidd,Gub}
\beq
     \ln Z_5(g)=S_{\rm CT}+S_{\rm QFT},
\eeq
where $S_{\rm QFT}$ represents the action for the boundary field theory,
and the explicit forms of these actions are given below.
One more important point is that
the boundary is pulled to the position of the brane
($y=0$). Then QFT in
$\ln Z_5(g)$ is replaced by a
cutoff theory since $y$ has the meaning of the energy scale of the QFT.
In this case, we expect extra counter terms
in $S^{\rm eff}_b$ due to the loop corrections
coming from the cut-off QFT which couples with the boundary metric.
It is shown below that this corrections play an impotant role.
The action, $S_{\rm CT}$, is divergent at the boundary ($y=-\infty$) 
since they have been introduced to cancel out the divergences
which appear there. However they are
finite at the brane position, and they are rewritten in terms of
the induced metric on the brane into the form of the part of 4d gravity
action.

\vspace{.3cm}
The alternative way to get $S^{\rm eff}_b$ is performing the path
integral directly in
the expression (\ref{effective}). We firstly consider this method
since we like to estimate $\lambda$ also for the case of $\Lambda>0$ where
there is no boundary in the bulk space.
The simplest estimation is given by the semi-classical approximation,
and it is obtained by substituting the classical 
solutions given in the previous 
section into $S_5$. Then the effective 4-dimensional action is obtained
by integrating over the fifth coordinate $y$. 

Although our solutions are given in the
form of (\ref{metrica}), we can write the following more general form
for the metric, 
\beq
 ds^2= A^2(y) g_{\mu\nu}(x)dx^{\mu}dx^{\nu}
           +dy^2  \, \label{metricag}.
\eeq
And the resultant effective 4-dimensional action is given as follows,
\beq
  S^{\rm eff}_b={1\over 2\kappa_4^2}
           \int d^4x \sqrt{-g}(R^{(4)}-2\Lambda_4+\cdots),  \label{4deff}
\eeq
\beq
  {1\over 2\kappa_4^2}={1\over 2\kappa^2}\int_0^{y_H} A^2dy  \label{plank}
\eeq
\beq
  \Lambda_4=4{\int_0^{y_H}dyA^2(AA''+{3\over 2}A'^2+{\Lambda\over 4}A^2)
            +\partial_yA|_{y=0}
              +{\kappa^2\tau\over 8}\over\int_0^{y_H} A^2dy}. \label{cosmo}
\eeq
Here, the dots in (\ref{4deff}) denote other 4-dimensional modes remained.

The results given by (\ref{4deff}) $\sim$ (\ref{cosmo}) are correct 
if we need no other terms coming from the loop corrections.
Such a case would be realized when the five dimensional theory
is constructed in a conformal invariant form, and the classical
solution is consistent with this invariance.
When the above action (\ref{4deff}) is correct, we should obtain
the same equation with the 4d part of the 5d equations, which is
given in the first equation in (\ref{Einstein2}). In $S^{\rm eff}_b$,
two parameters, $\kappa_4$ and $\Lambda_4$, are derived. The
corresponding parts are read as, $({\dot{a_0}\over a_0})^2=\lambda
+{2\kappa^4\tau\rho_b(t)\over 36}(1+{\rho_b(t)\over 2\tau})-{k\over a_0^2}$
from the first equation of (\ref{Einstein2}) by the replacement
\cite{BDEL},
$\tau\to \tau+\rho_b(t)$, where $\rho_b(t)$ denotes the matter density
on the brane. We should have taken into account of this term, $\rho_b(t)$,
at the starting
point, but there is no problem here since 
it is taken to be zero in our solution. 
The resultant correspondences are obtained as,
\beq
   3\lambda = \Lambda_4, \label{condicos1}
\eeq
\beq
   {\kappa^4\tau\over 3} = \kappa_4^2. \label{condicos2}
\eeq
The second relation 
(\ref{condicos2}) however should be remained as a useful one.

\vspace{.3cm}
Our next task is to see the consistency of the above two relations
by using (\ref{plank}) and (\ref{cosmo}) for our solution of $\lambda>0$.
After a little calculation, we can see that the first relation 
(\ref{condicos1}) is satisfied for both our solutions of $\Lambda>0$
and $\Lambda<0$. In other words, no extra constraint is needed from
(\ref{condicos1}). However the second condition is non-trivial, and we
obtain a new constraint on the parameter.

For the solution of $\Lambda<0$, (\ref{condicos2}) can be written as
\beq
 {1\over \sqrt{1+x}}=\sqrt{1+x}-x\ln({1+\sqrt{1+x}\over \sqrt{x}}), 
            \label{lapo}
\eeq
where $x=\lambda/\mu^2$. This equation has only one solution, $x=0$ or
$\lambda=0$. At a glance,
this result seems to be inconsistent since we 
have used the solution for $\lambda>0$ in this analysis. However we should
notice the following points. (i)
The solution used covers the one of $\lambda=0$ as a limit, and (ii)
the above $S^{\rm eff}_b$ in (\ref{4deff}) is obtained without including any
loop-correction. 
We could interpret this result, $\lambda=0$,
as the reflection of the conformal
invariance of the bulk theory since the second point mentioned above
is reasonable when the loop-corrections are cancelled out. 
In other words, the Poincare invariant solution (RS solution)
can be obtained by imposing the conformal invariance
of the theory not by the fine-tuning of the parameters.

\vspace{.3cm}
Therefore, in order to obtain a non-zero $\lambda$
from (\ref{condicos2}), we must include quantum corrections in deriving 
$S^{\rm eff}_b$ through the
path-integral over $G_{\mu\nu}$ and other fields.
They modify the form of (\ref{lapo}).
To consider in this way is quite natural since there is no reason to consider
the conformal invariance of the bulk theory for the background with small
$\lambda$. 

When
the conformal invariance is slightly broken, the corrections 
would appear generally in the forms of cosmological, Einstein
terms and other general coordinate invariant forms,
\beq
  \Delta S= {1\over 2\kappa^2}
      \int d^4x\sqrt{-g}(\epsilon_2(\lambda)/\mu R^{(4)}+8\epsilon_0(\lambda)\mu+\cdots),  
                  \label{Qcorr}
\eeq
where $\epsilon_2(\lambda)$ and $\epsilon_0(\lambda)$ are the dimensionless correction
terms scaled by the physical dimension-full parameter $\mu$. The other
corrections are denoted by $\cdots$.

We did't perform an explicit calculation of the loop corrections, 
but the corrections could be obtained as
functions of $A(y)$. And they can be integrated over $y$ as in the case of 
tree approximation. In general, the corrections are 
also dependent of the parameters
contained in $A(y)$, so they are expressed as $\epsilon_2(\lambda)$ and 
$\epsilon_0(\lambda)$. In this sense, 
it is impossible to see the precise constraint on $\lambda$
from (\ref{condicos2}) without knowing the explicit form of $\epsilon_2(\lambda)$ 
and $\epsilon_0(\lambda)$. 
However, we can see the possibility that the quantum corrections in the
bulk theory could give the small $\lambda$. For a while we consider 
$\epsilon_2$ as a small constant, then it appears on the right hand side
of (\ref{lapo}) as 
$\sqrt{1+x}-x\ln({1+\sqrt{1+x}\over \sqrt{x}})+\epsilon_2$. And we find
a non-trivial solution in this case by solving the approximated equation,
\beq
  x\ln x+4\epsilon_2=0,
\eeq
for small $x$. We also obtain a non-trivial solution from (\ref{condicos1})
when the quantum corrections are included, and a relation of the corrections
would be obtained. 

In the following,
we can see these points by using an alternative formulation
where the correction terms are independent of the parameter $\lambda$.
The formulation is based on the AdS/CFT correspondence, and
$\lambda$ 
appears as a result of conformal symmetry breaking in the quantum 
field theory on the boundary.
Before showing it, we examine the case of $\Lambda>0$.

\vspace{.3cm}
In the case of $\Lambda>0$, the bulk space has no 
spacial boundary as in the
case of negative $\Lambda$, so we can not expect 
the correspondence of the gravity
and the field theory on the boundary. But we can calculate $S_b^{\rm eff}$
as above and study the solution of (\ref{condicos2}) as above.
Using the explicit form of the solution, 
(\ref{condicos2}) leads to the following equation,
\beq
 {1\over \sqrt{x_d-1}}=x_d\sin^{-1}({1\over\sqrt{x_d}})-\sqrt{x_d-1}, 
            \label{lapo2}
\eeq
where $x_d=\lambda/\mu_d^2$. This equation has no solution for
$x_d\geq 1$, which is the allowed value of $x_d$.
Then quantum corrections are essential to get a non-trivial solution
in this case. In fact, there is no supersymmetry in $dS_5$ \cite{Witt},
so there is no conformal symmetry.
When we consider the corrections given above, the equation
(\ref{lapo2}) is modified by adding  $\epsilon_2$ on the
right hand side. Assuming that $\epsilon_2$ is independent of $x_d$
and small, we obtain the solution $\sqrt{x_d}=1/(3\epsilon_2)$. Then
it seems to be possible to have a reasonable solution for $\Lambda>0$
when quantum corrections are taken into account. However we can see that
this result is inconsistent with the analysis given in the next section.

\vspace{.3cm}
We now return to the case of $\Lambda<0$, where
the effective 4-dimensional action, $S^{\rm eff}_b$,
can also be obtained from the holographic viewpoint 
mentioned above. According to the formulation given above,
we can write the effective action as
\beq
  S^{\rm eff}_b = S_{\rm b2}+S_{\rm QFT}.
\eeq
Here, the QFT part is given as
\beq
  S_{\rm QFT}=\int d^4x( L_{\rm QFT}+\Sigma_i\lambda_iO^i), \label{qft}
\eeq
where $O^i$ are the composite operators of the fields contained 
in $L_{\rm QFT}$
and $\lambda_i$ are their corresponding
sources which are given as the boundary values
of the bulk fields at the brane position.
And the first term is defined as
\beq
   S_{\rm b2}={1\over 2}S_b+S_{\rm CT},
\eeq 
where $S_{\rm CT}$ denotes the counter term stated above.
This counter term has been obtained for the metric given in the form 
\cite{HeSk,BaKr},
\beq
 ds^2= {L^2\over \rho^2}\left\{g_{\mu\nu}(x,\rho)dx^{\mu}dx^{\nu}
                +d\rho^2\right\}
             \, \label{metricarho},
\eeq
where $g_{\mu\nu}(x,\rho)$ is expanded near the boundary ($\rho =0$)
in the series of $\rho$ as
\beq
  g_{\mu\nu}(x,\rho)=g_{\mu\nu}^{(0)}+g_{\mu\nu}^{(2)}\rho^2 + \cdots,
\eeq
where the higher order terms are denoted by $\cdots$.
Our solution of $\Lambda<0$ is written in this form by taking as
\beq
   L={1\over\mu}, \qquad g_{\mu\nu}(x,\rho)=
         (1-{\lambda\over 4}\rho^2)^2 g_{\mu\nu}(x)  \label{counter}
\eeq
\beq
  \rho={2\over \sqrt{\lambda}}{\rm tanh}({\sqrt{\lambda} z\over 2}),
\eeq
where
$z=\rm{sgn}(y)(\lambda)^{-1/2}\ln(\rm{coth}[\mu(y_H-|y|)/2])$ and
the position of the brane $z_0$ ($y=0$) is given by
\beq
 z_0={1\over\sqrt{\lambda}}\rm{arcsinh}({\sqrt{\lambda}\over \mu}).
\eeq
Hence we find
$g_{\mu\nu}^{(0)}=g_{\mu\nu}(x)$,
$g_{\mu\nu}^{(2)}=-{\lambda\over 2}g_{\mu\nu}^{(0)}$, and
$g_{\mu\nu}^{(4)}={\lambda^2\over 16}g_{\mu\nu}^{(0)}$.
This is consistent with the results given in \cite{HeSk,BaKr} when
we take the four dimensional Riemann tensor in the de Sitter form, 
$R_{\mu\nu\lambda\sigma}=-\lambda(g_{\mu\lambda}g_{\nu\sigma}
     -g_{\mu\sigma}g_{\nu\lambda})$.
So we obtain the same form of
counter terms with the one given in \cite{HeSk,BaKr}. 

The counter terms are written by the induced metric on the brane.
It is easily seen that the induced metric on the brane (at $z=z_0$)
is equal to $g_{\mu\nu}(x)$, i.e.
\beq
  {L^2\over \rho^2}g_{\mu\nu}(x,\rho)|_{z=z_0}=g_{\mu\nu}(x).
\eeq
This is independent on $\lambda$, and we obtain
\beq
    S_{\rm b2} = \int dx^4\sqrt{-g}\Bigg\{L_{\rm brane}
              -({\tau\over 2}+b_0)
     -b_2R-b_4R_2\Bigg\}, \label{baction2}
\eeq
where $b_0=-(6/L)/(2\kappa^2)$, $b_2=-(L/2)/(2\kappa^2)$, 
$b_4=2L^3/(2\kappa^2)$ and
\beq
  R_2 = -{1\over 8} R_{\mu\nu} R^{\mu\nu} + {1\over24} R^2\ .
\eeq
From the above result, we obtain 
\beq
  {1\over 2\kappa_4^2}=-b_2 , 
     \qquad -2\Lambda_4={b_0+\tau/2\over b_2}.
\eeq
Then we obtain the same result from both (\ref{condicos1}) and 
(\ref{condicos2}), 
\beq
  {\kappa^2\tau\over 6}=\mu ,
\eeq
where $\mu=1/L$. Hence, we find $\lambda=0$ again. This result is consistent
with the analysis given above in the case of the 
conformal invariant bulk theory. In this case, QFT is equivalent to CFT
in the sense that any quantum corrections to $b_0$ and $b_2$
parts are not added. Even if we consider CFT, $b_4$ should be modified
by the anomaly term coming from the loop corrections in CFT. There would
be a possibility to find 4d de Sitter solution within CFT by considering
the higher derivative gravity with the anomaly \cite{NO}.

As in the previous analysis, we should consider QFT, in which conformal
symmetry is slightly broken, to find a solution with small $\lambda$. Then
we include the quantum corrections in the effective four-dimensional
action, especially for $b_0$ and $b_2$, as discussed above.
Then we should modify $b_0$ and $b_2$ as follows,
\beq
 b_0=-(6/L)/(2\kappa^2)+\bar{\epsilon}_0\mu^4 \, ,
   \qquad   b_2=-(L/2)/(2\kappa^2)+\bar{\epsilon}_2\mu^2 \, .
\eeq
where $\bar{\epsilon}_0$ and $\bar{\epsilon}_2$ represent the dimensionless
loop-correction terms derived from $S_{\rm QFT}$. In this case,
$\bar{\epsilon}_0$ and $\bar{\epsilon}_2$ are independent on $A(y)$, then
$\lambda$ can be estimated by using (\ref{condicos1}) and (\ref{condicos2})
in terms of $\bar{\epsilon}_0$ and $\bar{\epsilon}_2$.
And we obtain
\beq
   \lambda=\bar{\epsilon}{\mu^3\over M^3}\mu^2,
\eeq
where $1/2\kappa^2=M^3$. The coefficient $\bar{\epsilon}$ is given as 
$\bar{\epsilon}=4\bar{\epsilon}_2$
and $\bar{\epsilon}=\bar{\epsilon}_0/6$ from (\ref{condicos1}) and 
(\ref{condicos2}) respectively. Then the following relation of the corrections,
\beq
   \bar{\epsilon}_0=24\bar{\epsilon}_2,  \label{correc2}
\eeq
is needed from the consistency. 
Hence a small
$\lambda$ is obtained as a quantum correction from the boundary QFT.
Then, the conformal symmetry should be broken slightly in the boundary
QFT. 

This implies that the gravity on the bulk manifold, which is described
by our solutions with a small $\lambda$, would describe a non-conformal 
quantum field theory on the boundary. It will be an interesting 
problem to see
what kind of field theory can be seen on the boundary. 
The relation (\ref{correc2}) given above could be a clue to this problem.

@
\section{Observational constraint on bulk space}

In this section, the 5d bulk space
is constrained from observational information.
The cosmological constant $\lambda$ and the Planck mass $M_{\rm pl}$
are set to the measured values,
$\lambda_{obs} \sim 10^{-122}M_{\rm pl}^2$ and
$M_{\rm pl} \sim 10^{19} {\rm GeV}$. As a merit of this setting,
quantum corrections to these are taken into account implicitly.
Especially, quantum corrections to
$\lambda$ are essential, as mentioned in the previous section.
Inserting Eq. (\ref{condicos2}) into Eq.(\ref{4cos1}) leads to
$\Lambda=6\lambda_{obs}-3M^6/(2M_{\rm pl}^4) < - 10^{-90}M_{\rm pl}^2$,
because of $1/2\kappa^2=M^3$, $1/2\kappa_4^2=M_{\rm pl}^2$.
The upper limit of $\Lambda$ is determined by the observational constraint
$M > 10^4 {\rm GeV}$, which comes from the condition that in the effective
4d Friedmann equation derived from the 5d theory the $\rho_b$ term
should be larger than the $\rho_b^2$ term at the epoch of Big Bang
Nucleosynthesis \cite{Cline,Yahiro}. The upper limit is
negative, and its absolute value is quite small.
This indicates that
the $dS_5$ space is prohibited,
while the $AdS_5$ space is allowed for almost
all negative $\Lambda$.
In the previous section, it is found that there exists a solution
$\lambda=0$
for $\Lambda<0$ but none for $\Lambda>0$, when the classical limit is taken.
Quantum corrections to $\lambda$ and $M_{\rm pl}$ relax the situation
so that a small positive $\lambda$ can exist
for each of $\Lambda>0$ and $\Lambda<0$.
The possibility of the small $\lambda$ for the case of $\Lambda>0$, however,
is excluded by the observational constraint on $\Lambda$.

It is possible to make a similar analysis with Eq. (\ref{plank})
instead of Eq.(\ref{condicos2}), since both are considered to be identical.
The two analyses should give a consistent constraint on $\Lambda$.
This consistency is confirmed, as follows.
When $\Lambda >0$, the solution (\ref{desit}) is
inserted into Eq. (\ref{plank}).
This leads to a condition for
$\alpha=1/\sqrt{x_d}=\mu_d/\sqrt{\lambda_{obs}}$,
$2\sqrt{\lambda_{obs}}M_{\rm pl}/M^3
 =\{ \arcsin{(\alpha)}-\alpha \sqrt{1-\alpha^2} \}/\alpha^3$.
The left hand side contains quantum corrections
implicitly. The left hand side is smaller than $10^{-16}$ because of
$M > 10^4 {\rm GeV}$, but the right hand side is larger than $2/3$
in the range $0 \leq \alpha \leq 1$ determined from
Eq.(\ref{4cos1}).
Hence, no $\alpha$ satisfies the equation, as expected above.

Now we consider quantum corrections to the right hand side by adding $\Delta
S$ defined in Eq. (\ref{Qcorr}) to $S_b^{\rm eff}$.  The resultant equation
is
the same as the equation shown above, except that a correction term,
$\epsilon_2(\lambda)/\alpha$, is added to the right hand side.
As an important property, the correction term is positive, as shown
in the previous section. The right hand side is still larger than $2/3$.
Therefore, no $\alpha$ satisfies the equation.
The $dS_5$ bulk space is thus prohibited for the small observable
cosmological constant.

Also in the case of $\Lambda < 0$, the corresponding condition is obtained
for $\alpha=1/\sqrt{x}=\mu/\sqrt{\lambda_{obs}}$
by inserting the solution (\ref{metrica4}) into
Eq. (\ref{plank}) and considering the quantum correction. The condition is
$2\sqrt{\lambda_{obs}}M_{\rm pl}^2/M^3
  =\{-{\rm arcsinh}(\alpha)+\alpha \sqrt{1+\alpha^2}\}/\alpha^3
  +\epsilon_2(\lambda)/\alpha.$
In the classical limit, where $\epsilon_2(\lambda)=0$,
this equation is satisfied when $\alpha > 10^{16}$, because the left hand
side is less than $10^{-16}$. The allowed range of $\alpha$ corresponds
to $x < 10^{-32}$ or
$|\Lambda| > 10^{32} \lambda_{obs} \sim 10^{-90}M_{\rm pl}^2 $.
The quantum correction makes
the lower limit of $|\Lambda|$ go up, since it is positive.
But the shift is quite small, because so is $\epsilon_2(\lambda)$.
The allowed range of $\Lambda$ is consistent with the one mentioned above.


\section{Summary and Cosmological Implications}

From the viewpoint of 5-dimensional brane-world, we have examined
the 4-dimensional universe with a small cosmological constant.
The three brane considered here has been embedded in $AdS_5$, and it can be
considered as an extended
part near the horizon of the configuration realized by the stack of 
D3 branes in the type IIB
superstring theory. In this sense, supersymmetry could be preserved in the bulk
and other properties of the superstring theory would be expected. 
Then we can expect that
the quantum corrections for such a configuration would be cancelled out when
the conformal symmetry remains.
In this case, we could arrive at our conclusion of $\lambda=0$, which has been
actually derived without any quantum corrections comming from the bulk theory.
This result would be important in the sence that any finite $\lambda$ should be
forbidden by the conformal invariance in the bulk and we
would need not any fine-tuning for the Poincare invariance
in the 4d space-time on the brane. 

Hence, some symmetry breaking is expected for producing the 
quantum corrections given above in order to get a small $\lambda$. 
In fact, our solutions used here are
deformed from the
$AdS_5$ by the metric of three space, $g_{ij}=a_0(t)^2\gamma_{ij}$, due to
non-zero $\lambda$. The supersymmetric domain wall solutions are known
in 5d supergravity by considering flux condensation in the form of $AdS_5$.
Then, it would be an intersting work to see whether some
supersymmetries are preserved or not for our solutions in the bulk 5d
supergravity. It will be remained as a future work 
to see a possible solution with some small conformal symmetry breaking
and to estimate
$\lambda$ quantitatively in terms of such a solutions derived from 
more concrete theory.

In any case,
the universe with small $\lambda$ would be explained
by assuming a five dimensional
theory with a weak breaking of the conformal symmetry. 
This symmetry breaking
can be considered as a reflection of the deformation of AdS
space as mentioned above. 
In the context of AdS/CFT
correspondence, this deformation would break also the conformal
invariance of the field theory on the boundary (QFT). In fact, we could show
that a small cosmological constant on the brane is obtained by
considering quantum corrections coming from
the QFT which couples with the gravity on the brane in the framework
of holography. 
The cosmological constant on the brane and
the conformal symmetry breaking in the bulk theory or in QFT
are related intimately to each other.

As a natural statement, we can say that this small cosmological constant
causes the observed acceleration in the present universe.
Its amount is controlled by the symmetry
not on the brain but in the bulk.
For the brane-world with observable cosmological constant,
the $dS_5$ bulk space seems to be prohibited because of
the observational constraint on $M_5$.
In contrast,
the $AdS_5$ bulk space is allowed for almost all negative $\Lambda$.
It is thus highly expected that our universe is embedded not in $dS_5$ but
in $AdS_5$.

While, some amount of dark matter is also expected from the recent
observations.
This would be also explainable
from the brane-world viewpoint. From our analysis, which would be reported
soon in a separate paper, the localization of massive scalar
with mass smaller than $3\sqrt{\lambda}/2$ in the bulk would be
trapped on the brane and this scalar would interact with matters on the
brane only through gravitation. So this scalar could be considered as
a candidate for the cold dark matter.
As an important fact, it should be stressed that the phenomena occur only
when
the positive cosmological constant exists.

\vspace{.3cm}
\noindent {\bf Acknowledgements}\par
Both authors are supported by the grants of the Ministry of 
Education, Science, Sports and Culture of Japan in completing this work. 


\end{document}